\documentclass[12pt]{article}
\makeatletter
\newcommand{\singlespacing}{\let\CS=\@currsize\renewcommand{\baselinestretch}{1}\tiny\CS}
\usepackage{epsfig}

\usepackage{graphicx}
\usepackage{verbatim}   
\usepackage{color}      
\usepackage{subfigure}  
\usepackage{amsfonts}
\usepackage{subfigure}  
\usepackage[normalem]{ulem}
\begin{document}
\baselineskip=24pt
\parskip = 10pt
\def \qed {\hfill \vrule height7pt width 5pt depth 0pt}
\newcommand{\ve}[1]{\mbox{\boldmath$#1$}}
\newcommand{\IR}{\mbox{$I\!\!R$}}
\newcommand{\1}{\Rightarrow}
\newcommand{\bs}{\baselineskip}
\newcommand{\esp}{\end{sloppypar}}
\newcommand{\be}{\begin{equation}}
\newcommand{\ee}{\end{equation}}
\newcommand{\beanno}{\begin{eqnarray*}}
\newcommand{\inp}[2]{\left( {#1} ,\,{#2} \right)}
\newcommand{\eeanno}{\end{eqnarray*}}
\newcommand{\bea}{\begin{eqnarray}}
\newcommand{\eea}{\end{eqnarray}}
\newcommand{\ba}{\begin{array}}
\newcommand{\ea}{\end{array}}
\newcommand{\nno}{\nonumber}
\newcommand{\dou}{\partial}
\newcommand{\bc}{\begin{center}}
\newcommand{\ec}{\end{center}}
\newcommand{\2}{\subseteq}
\newcommand{\cl}{\centerline}
\newcommand{\ds}{\displaystyle}
\newcommand{\mr}{\mathbb{R}}
\newcommand{\mn}{\mathbb{N}}
\def\refhg{\hangindent=20pt\hangafter=1}
\def\refmark{\par\vskip 2.50mm\noindent\refhg}

\title{\sc Stationary GE-Process and its Application in Analyzing Gold Price Data}

\author{\sc Debasis Kundu \footnote{Department of Mathematics and Statistics, Indian Institute of
Technology Kanpur, Pin 208016, India.  E-mail:
kundu@iitk.ac.in, Phone no. 91-512-2597141, Fax no. 91-512-2597500.}}

\date{}
\maketitle

\begin{abstract}

  In this paper we introduce a new discrete time and continuous state space stationary process $\{X_n; n = 1, 2,
  \ldots \}$, such that $X_n$ follows a two-parameter generalized exponential (GE) distribution.  Joint distribution
  functions, characterization and some dependency properties
  of this new process have been investigated.  The GE-process has three unknown parameters, two shape parameters and
  one scale parameter, and due to this reason it is more flexible than the existing exponential process.   In
  presence of the scale parameter, if the
  two shape parameters are equal, then the maximum likelihood estimators of the unknown parameters can be obtained
  by solving one non-linear equation and if the two shape parameters are arbitrary, then the maximum likelihood
  estimators can be obtained by solving a two dimensional optimization problem.  Two {\color{black}
    synthetic} data sets, and
  one real gold-price data set have been analyzed to see the performance of the proposed model in
  practice.   Finally some generalizations have been indicated.

\end{abstract}




\section{\sc Introduction}

Gaussian assumptions are quite common in the theoretical development of any Markovian process.  Very few Markovian
models have been developed with out the Gaussian assumptions.  If the data indicate any non-Gaussian behavior,
the usual method is to attempt to remove the skewness of the data by taking suitable transformation, and then use
the Gaussian process to the transformed data.  Although, it has been criticized severely in the literature.  Nelson
\cite{Nelson:1976} as well as Granger and Andersen \cite{GA:1978} correctly pointed out that quite often the transformed
economic data are no where near Gaussian.  It may not be very surprising, because Weiss \cite{Weiss:1975} showed that
if $\{X_t\}$ is a stationary process, and $f(\cdot)$ is a one-to-one function, then {\color{black}$Y_t = f(X_t)$} is time reversible
if and only if $\{X_t\}$ is time reversible.  Therefore, it is immediate that a process cannot be transformed to a time
reversible Gaussian process unless the process itself is time reversible.

Due to this reason, several non-Gaussian processes have been introduced and studied quite extensively in the literature.
For example, stationary exponential process by Tavares \cite{Tavares:1980}, Weibull and gamma processes by Sim \cite{Sim:1986}
Logistic process by Arnold \cite{Arnold:1993}, Pareto process by Arnold and Hallet \cite{AH:1989}, see also
Arnold \cite{Arnold:2001}, semi-Pareto process by Pillai \cite{Pillai:1991}, Marshall-Olkin bivariate Weibull processes by
Jose, Risti{\'c} and Joseph \cite{JRJ:2011}, generalized Weibull process by Jayakumar and Girish Babu
\cite{JG:2015} and see the references cited therein.  In all these cases the emphasis is to develop a stationary
process which has specific marginals.  In most of the cases they have been developed using minification process
of autoregressive sequences.

Recently, generalized exponential (GE) distribution has received a considerable amount of attention in the
statistical literature.  It is a positively skewed distribution, and it can be used quite effectively to analyze
lifetime data as an alternative to the popular Weibull or gamma distributions.  The aim of this paper is to develop
a stationary process whose marginals are identically distributed GE distributions.  The GE process has been obtained
using the maxification process of moving average (MA) sequences.  If we use the $q$-th order MA process, then
a $q$-dependent sequence with GE marginals can be generated.  It has been obtained quite naturally by using the
property that the GE distribution is closed under maximization.

We study different properties of the GE process $\{X_n; n = 1,2, \ldots\}$, when $q$ = 1.  The joint, marginal and
conditional distributions have been obtained.  Some characterizations and a mixture representation have been provided.  The
generation of the GE process is quite straight forward, hence simulation experiments can be performed quite
conveniently.  The distributions of the maximum and minimum of the GE process and also the probability mass function
of the stopping time have been presented. The GE process has two shape parameters and one scale parameter.  When the
two shape parameters are equal, then the joint distribution of $X_n$ and $X_{n+1}$ has a very convenient copula structure, hence several dependency properties and also dependency measures can be easily obtained.  We have
provided some dependency properties of the proposed GE process, and also provided different dependency measures
of the process.

The estimation of the unknown parameters is an important issue in a real data analysis problem.  The GE process
has three unknown parameters.  The most natural estimators will be the maximum likelihood estimators.  If
the two shape parameters are same, the maximum likelihood estimators (MLEs) can be obtained by solving one non-linear
equation.  Although, we could not prove it theoretically, it is observed from the profile likelihood function
plot, that the MLEs exist and they are unique.  If the two shape parameters are not equal, the maximum likelihood estimators can be obtained by solving a two dimensional optimization problem.  In this case also, from the contour
plot, it is observed that the MLEs exist and they are unique.  We have analyzed {\color{black} two synthetic data sets, 
and one gold-price data set} to show how the proposed model behaves in practice.  Finally we propose
some generalizations and open problems.

The main contribution of this paper is to introduce stationary GE process and derive several of its properties.
Although, Weibull and gamma processes have been discussed in the literature quite extensively, the same is not
true in case of GE process, although, GE distribution becomes very popular in the last two decades.  Another
important contribution is the estimation of the unknown parameters of the proposed GE process.  Although, Weibull
and GE processes have been discussed quite extensively, no where the estimation procedures have been proposed.
Similar estimation  procedures what we have used here, can be used for Weibull and {\color{black} gamma} processes also.

The rest of the paper is organized as follows.  In Section 2, we briefly describe the GE distribution.  The
stationary GE process is proposed in Section 3 and its several properties have been presented.  The maximum
likelihood estimators are described in Section 4.  The analyses of {\color{black} two synthetic data sets
  and one gold price  data set are presented in Section 5 and Section 6, respectively}.
Some generalizations and open problems are indicated in Section 7.

\section {\sc GE Distribution: A Brief Review}

The generalized exponential distribution was originally introduced by Gupta and Kundu \cite{GK:1999} as a special
case of the exponentiated Weibull distribution of Mudholkar and Srivastava \cite{MS:1993}.  The two-parameter
GE distribution has the following cumulative distribution function (CDF);
\be
F_{GE}(t; \alpha,\lambda) = \left (1 - e^{-\lambda t} \right )^{\alpha}; \ \ \  t > 0,   \label{cdf-ge}
\ee
and 0, otherwise.  Here, $\alpha$ and $\lambda$ are the shape and scale parameters, respectively.  The corresponding
probability density function (PDF) becomes;
\be
f_{GE}(t; \alpha,\lambda) = \alpha \lambda e^{-\lambda t} \left (1 - e^{-\lambda t} \right )^{\alpha-1}; \ \ \  t > 0,
\label{pdf-ge}
\ee
and 0, otherwise.  A GE random variable with the CDF (\ref{cdf-ge}) and PDF (\ref{pdf-ge}) will be denoted by
GE$(\alpha,\lambda)$, and if $\lambda$ = 1, it will be denoted by GE$(\alpha)$.  For a GE$(\alpha)$ random
variable the corresponding PDF and CDF will be denoted by $f_{GE}(t;\alpha)$ and $F_{GE}(t; \alpha)$, respectively.

It is immediate that when $\alpha$ = 1, the GE distribution becomes an exponential distribution.  Hence, the GE
distribution is an extension of the exponential distribution, similar to the Weibull and gamma distributions but
in different ways.  It has been observed that the shapes of the PDF and hazard functions of a GE distributions are quite similar to the Weibull and gamma distributions.  The hazard function of a GE distribution can be an increasing,
decreasing or constant depending on the shape parameter.  Since the CDF of a GE distribution is in compact form,
hence, the generation of a random sample from a GE distribution is quite straight forward.  The GE distribution is
closed under maximum and it can be used quite effectively in place of gamma or Weibull distribution for data analysis
purposes.

Different moments, order statistics, record values, various estimation procedures, closeness with other distributions
like Weibull, gamma, log-normal, have been investigated by several authors.  It is observed that the GE distribution
is close to a gamma distribution than to a Weibull distribution.  Interested readers are referred to the review
articles by Gupta and Kundu \cite{GK:2007}, Nadarajah \cite{Nadarajah:2011},
Al-Hussaini and Ahsanullah \cite{AA:2015} and see the references cited therein.

\section{\sc GE Process and its Properties}

In this section first we define a stationary Markov process $\{X_n\}$, so that $X_n$ follows a GE distribution
and will investigate its several properties.

\noindent {\sc Definition:} Let $U_0, U_1, \ldots$ be a sequence of independent and identically distributed (i.i.d.)
Uniform $(0,1)$ random variables.  For $\alpha_0 > 0$ and $\alpha_1 > 0$, let us define a new sequence of random variables
\be
X_n = \max\{-\ln (1-U_n^{\frac{1}{\alpha_0}}),-\ln (1-U_{n-1}^{\frac{1}{\alpha_1}})\}.   \label{ge-process}
\ee
Then the sequence of random variables $\{X_n\}$ is called a GE process.

\noindent From the definition of the GE process, it is very easy to generate random samples from a stationary GE
process with a given $\alpha_0$ and $\alpha_1$.  We first generate random samples from $U(0,1)$, and then by the
required transformation, we can generate $\{X_n\}$.  The following Theorem provides the justification of the name
GE process.  {\color{black} It shows that the marginals follow GE distribution, and it is a stationary process}.

\noindent {\sc Theorem 1:} If the sequence of random variables $\{X_n\}$ is as defined in (\ref{ge-process}), then

\noindent (a) $\{X_n\}$ is a stationary Markov process.  \\
\noindent (b) $\{X_n\}$ follows GE$(\alpha_0+\alpha_1)$.

\noindent {\sc Proof:} Part (a) is trivial.

To prove part (b), note that
\beanno
P(X_n \le x) & = & P \left [ -\ln (1-U_n^{\frac{1}{\alpha_0}}) \le x,-\ln (1-U_{n-1}^{\frac{1}{\alpha_1}} \le x \right ] \\
& = & P \left [ U_n \le (1-e^{-x})^{\alpha_0},U_{n-1} \le (1-e^{-x})^{\alpha_1} \right ] \\
& = & \left (1-e^{-x} \right )^{\alpha_0+\alpha_1}.
\eeanno
\qed

\noindent The following result characterizes the GE process.

\noindent {\sc Theorem 2:} Suppose $X_1 \sim$ GE$(\alpha_0+\alpha_1)$ and $U_i$s are i.i.d. random variables with an
absolute continuous distribution function $F(x)$ on $(0,1)$.  Then the process as defined in (\ref{ge-process}) is
a strictly stationary Markov process if and only if $U_i$s are i.i.d. $U(0,1)$ random variables.

\noindent {\sc Proof:} `If' part is trivial.  To prove the `only if' part,let us assume that $F'(x) = f(x)$, for
$x > 0$.  Then from the definition of (\ref{ge-process}), we have for all $x \in (0, \infty)$,
\be
\left (1 - e^{-x} \right )^{\alpha_0+\alpha_1} = F((1 - e^{-x})^{\alpha_0})F((1 - e^{-x})^{\alpha_1}).   \label{1st-step}
\ee
Since, (\ref{1st-step}) is true for all $x \in (0, \infty)$, therefore, it can be written as
\be
y^{\alpha_0+\alpha_1} = F(y^{\alpha_0}) F(y^{\alpha_1}) \ \ \ \Leftrightarrow \ \ \frac{F(y^{\alpha_0})}{y^{\alpha_0}}
\times \frac{F(y^{\alpha_1})}{y^{\alpha_1}} = 1,
\ee
for all $y \in (0,1)$.  Therefore, for all $\alpha > 0$ and for all $y \in (0,1)$,
$$
\frac{F(y^{\alpha})}{y^{\alpha}} = 1,  \ \ \ \Rightarrow \ \ F(y) = y.
$$
\qed

\noindent The following result provides the joint distribution of $X_n$ and $X_{n+m}$, for $m \ge 1$.

\noindent {\sc Theorem 3:} If the sequence of random variables $\{X_n\}$ is defined as in (\ref{ge-process}), then the
joint distribution of $X_n$ and $X_{n+m}$, $F_{X_n,X_{n+m}} = P(X_n \le x, X_{n+m} \le y)$ is
\be
F_{X_n,X_{n+m}}(x,y) = \left \{ \begin{array}{lll}
  (1-e^{-x})^{\alpha_0+\alpha_1} (1-e^{-y})^{\alpha_0+\alpha_1} & \hbox{if} & m \ge 2  \\
  (1-e^{-x})^{\alpha_1} (1-e^{-y})^{\alpha_0} g(x,y) & \hbox{if} & m =1,
\end{array}
\right .
\ee
where $\ds g(x,y) = \min\{(1-e^{-x})^{\alpha_0},(1-e^{-y})^{\alpha_0}\}$.

\noindent {\sc Proof:} It mainly follows from the definition and considering the two cases $m$ = 1 and $m \ge 2$,
separately.  \qed

It immediately follows from Theorem 3 that $X_n$ and $X_{n+m}$ are independently distributed if $m > 1$, otherwise
they are dependent.  Now first we would like to study some dependency properties of $X_n$ and $X_{n+1}$.  The joint
distribution function of $X_n$ and $X_{n+1}$ can be written as
\be
F_{X_n,X_{n+1}}(x,y) = \left \{ \begin{array}{lll}
  (1-e^{-x})^{\alpha_0+\alpha_1} (1-e^{-y})^{\alpha_0} & \hbox{if} &   (1-e^{-x})^{\alpha_0} \le (1-e^{-y})^{\alpha_1}  \\
  (1-e^{-x})^{\alpha_1} (1-e^{-y})^{\alpha_0+\alpha_1} & \hbox{if} &   (1-e^{-x})^{\alpha_0} \ge (1-e^{-y})^{\alpha_1}.
\end{array}
\right .
\ee
Since $F_{X_n}(x) = (1-e^{-x})^{\alpha_0+\alpha_1}$ and $F_{X_{n+1}}(y) = (1-e^{-y})^{\alpha_0+\alpha_1}$, it is immediate
that for all $x > 0$, $y > 0$,
\be
F_{X_n,X_{n+1}}(x,y) \ge F_{X_n}(x) F_{X_{n+1}}(y).  \label{pqd}
\ee
Hence, $X_n$ and $X_{n+1}$ are positive quadrant dependent (PQD), therefore $Cov(X_n,X_{n+1}) > 0$.  It can be easily
verified from the definition that $X_n$ and $X_{n+1}$ has the total positivity of order two ($TP_2$) property.  Hence,
$(X_n,X_{n+1})$ has left tail decreasing (LTD) as well as left corner set decreasing (LCSD) properties, see for
example Nelsen \cite{Nelsen:2006}.

It can be easily seen that the joint distribution function $F_{X_n,X_{n+1}}(x,y)$ has the following copula function
\be
C(u,v) = \left \{ \begin{array}{lll}
  u v^{\frac{\alpha_0}{\alpha_0+\alpha_1}} & \hbox{if} & u^{\frac{\alpha_0}{\alpha_0+\alpha_1}} \le v^{\frac{\alpha_1}{\alpha_0+\alpha_1}} \\
  u^{\frac{\alpha_1}{\alpha_0+\alpha_1}} v & \hbox{if} & u^{\frac{\alpha_0}{\alpha_0+\alpha_1}} \ge v^{\frac{\alpha_1}{\alpha_0+\alpha_1}}.
\end{array}
\right .
\ee
Therefore, if we use $\ds \delta = \frac{\alpha_0}{\alpha_0+\alpha_1}$,then
\be
C(u,v) = \left \{ \begin{array}{lll}
  u v^{\delta} & \hbox{if} & u^{\delta} \le v^{1-\delta} \\
  u^{1-\delta} v & \hbox{if} & u^{\delta} \ge v^{1-\delta}.
\end{array}
\right .
\ee
Based on the copula function, the following dependence measures can be easily obtained.  For example, the  Kendall's
$\tau$ becomes
$$
\tau = \frac{\delta(1-\delta)(1-\delta(1-\delta))}{d^3+\delta(1-\delta)+\delta^2(1-\delta)^2+(1-\delta)^3}.
$$
It can be easily seen that the minimum value of Kendall's $\tau$ is zero, and it becomes maximum at
$\delta = \frac{1}{2}$, and the maximum value is $\frac{1}{3}$.  The Spearman's $\rho$ becomes
$$
\rho = \frac{3 \delta(1-\delta)}{\delta^2-\delta+2}.
$$
In this case the minimum value of Spearman's $\rho$ is zero, and it becomes maximum at 
$\delta = \frac{1}{2}$, and the maximum value is $\frac{3}{7}$.  Therefore, both Kendall's $\tau$ and Spearman's
$\rho$ become maximum when $\alpha_0 = \alpha_1$.

We need the following notations for further development.  The sets $S_1$, $S_2$ and the curve $C$ will be defined
as follows.
\bea
S_1 & = & \{(x,y); x > 0, y > 0, \left (1-e^{-x} \right )^{\alpha_0} < \left (1-e^{-y} \right )^{\alpha_1}\}  \label{s-1}\\
S_2 & = & \{(x,y); x > 0, y > 0, \left (1-e^{-x} \right )^{\alpha_0} > \left (1-e^{-y} \right )^{\alpha_1}\}  \label{s-2} \\
C & = & \{(x,y); x > 0, y > 0, \left (1-e^{-x} \right )^{\alpha_0} = \left (1-e^{-y} \right )^{\alpha_1}.\}  \label{c-curve}
\eea
Note that the curve $C$ has the parametric form $(t,\gamma(t))$, where $\ds \gamma(t) = -\ln \left (1 -
(1-e^{-t})^{\frac{\alpha_0}{\alpha_1}} \right )$, for $0 < t < \infty$.

{\color{black} The following theorem shows that the joint distribution of $X_n$ and $X_{n+1}$ is a singular
  distributions.  It means, although both $X_n$
  and $X_{n+1}$ are absolutely continuous, there is a positive probability that $X_n = X_{n+1}$.  The joint distribution
  function of $X_n$ and $X_{n+1}$ can be decomposed uniquely as an absolute continuous part and a singular part.}

\noindent {\sc Theorem 4:} If the sequence of random variables $\{X_n\}$ is same as defined in (\ref{ge-process}),
then the joint CDF of $X_n$ and $X_{n+1}$ can be written as
\be
F_{X_n,X_{n+1}}(x,y) = p F_a(x,y) + (1-p) F_x(x,y),
\ee
here
$$
p = \frac{\alpha_0^2 + \alpha_1^2}{\alpha_0^2+\alpha_1^2+\alpha_0 \alpha_1}, \ \ \ \ \
F_s(x,y) = (g(x,y))^{\frac{\alpha_0^2+\alpha_1^2+\alpha_0 \alpha_1}{\alpha_0 \alpha_1}} \ \ \ \ \hbox{and}
$$
$$
F_a(x,y) = \frac{\alpha_0^2+\alpha_1^2+\alpha_0 \alpha_1}{\alpha_0^2+\alpha_1^2} \left \{(1-e^{-x})^{\alpha_1}
(1-e^{-y})^{\alpha_0} g(x,y) \right \} - \frac{\alpha_0 \alpha_1}{\alpha_0^2+\alpha_1^2}
(g(x,y))^{\frac{\alpha_0^2+\alpha_1^2+\alpha_0 \alpha_1}{\alpha_0 \alpha_1}}
$$
\noindent {\sc Proof:} Note that $p$ and $F_a(x,y)$ can be obtained from $F_{X_n,X_{n+1}}(x,y)$ as follows
$$
p = \int_0^{\infty} \int_0^{\infty} \frac{\partial^2}{\partial x \partial y} F_{X_n,X_{n+1}}(x,y)
$$
and
$$
pF_a(x,y) = \int_0^x \int_0^y \frac{\partial^2}{\partial u \partial v} F_{X_n,X_{n+1}}(u,v).
$$
From
$$
\frac{\partial^2}{\partial x \partial y} F_{X_n,X_{n+1}}(x,y) = \left \{ \begin{array}{ccc}
  f_1(x,y) & \hbox{if} & (x,y) \in S_1   \\
  f_2(x,y) & \hbox{if} & (x,y) \in S_2,
\end{array}
\right .
$$
where
\beanno
f_1(x,y) & = & \alpha_0 (\alpha_0+\alpha_1)e^{-(x+y)}(1-e^{-x})^{\alpha_0+\alpha_1-1} (1-e^{-y})^{\alpha_0-1}  \\
f_2(x,y) & = & \alpha_1 (\alpha_0+\alpha_1)e^{-(x+y)}(1-e^{-x})^{\alpha_0-1} (1-e^{-y})^{\alpha_0+\alpha_1-1}, 
\eeanno
the expressions for $p$ and $F_a(x,y)$ can be obtained.  Once, we obtain $p$ and $F_a(x,y)$, $F_s(x,y)$ can be
obtained by subtraction.

Alternatively, the probabilistic arguments also can be given.  Suppose $A$ is the following event:
$$
A = \left \{-\ln \left (1 - U_n^{\frac{1}{\alpha_0}} \right ) > -\ln \left (1 - U_{n-1}^{\frac{1}{\alpha_1}} \right )
\right \} \cap  \left \{-\ln \left (1 - U_n^{\frac{1}{\alpha_1}} \right ) > -\ln \left (1 - U_{n+1}^{\frac{1}{\alpha_0}},
\right ) \right \}
$$
then
$$
P(X_n \le x, X_{n+1} \le y) = P(X_n \le x, X_{n+1} \le y|A)P(A) + P(X_n \le x, X_{n+1} \le y|A^c)P(A^c).
$$
Now consider
$$
P(X_n \le x, X_{n+1} \le y|A) = (g(x,y))^{\frac{\alpha_0^2+\alpha_1^2+\alpha_0 \alpha_1}{\alpha_0 \alpha_1}}
$$
and
$$
P(A) = P \left [ U_{n-1}^{\frac{\alpha_0}{\alpha_1}} < U_n,U_{n+1}^{\frac{\alpha_1}{\alpha_0}} < U_n \right ]
= \int_0^1 u^{\frac{\alpha_1}{\alpha_0}+\frac{\alpha_0}{\alpha_1}}du = \frac{\alpha_0 \alpha_1}{\alpha_0^2+\alpha_1^2+\alpha_0 \alpha_1} = 1-p.
$$
Moreover,  $P(X_n \le x, X_{n+1} \le y|A^c)$ can be obtained by subtraction.  Clearly, $F_s(x,y)$ is the singular part,
as its mixed partial derivative is 0 in $S_1 \cup S_2$, and $P(X_n \le x, X_{n+1} \le y|A^c)$ is the absolute
continuous part, as its mixed partial derivative is a proper bivariate density function.   \qed

Now we would like to obtain the joint probability density function with respect to a proper dominating measure.  It
will be needed to compute the maximum likelihood estimators of the unknown parameters and other associated statistical
inferences based on density functions.  We consider the following dominating measure, similarly as in Bemis, Bain
and Higgins \cite{BBH:1972}.  The dominating measure is the two dimensional usual Lebesgue measure on $S_1 \cup S_2$,
and one dimensional Lebesgue measure defined on the curve $C$.  Here a length is defined as the arc length on the
curve $C$.  One natural question is whether we can get different results using different dominating measures.
Fortunately, the answer is negative due to the application of the elementary results by Halmos \cite{Halmos:1950},
see also Bemis, Bain and Higgins \cite{BBH:1972} in this connection.

{\color{black} The following theorem provides the explicit form of the joint PDF of $X_n$ and $X_{n+1}$
  based on the above dominating measure}.  

\noindent {\color{black}{\sc Theorem 5:}} If $\{X_n\}$ is same as defined in Theorem 3, then the joint PDF of $X_n$ and $X_{n+1}$
for $x > 0$ and $y > 0$ is
\be
f_{X_n,X_{n+1}}(x,y) = \left \{ \begin{array}{ccc}
  f_1(x,y) & \hbox{if} & (x,y) \in S_1 \\
  f_2(x,y) & \hbox{if} & (x,y) \in S_2 \\
  f_0(x) & \hbox{if} & y = \gamma(x),
\end{array}
\right .   \label{joint-pdf}
\ee
where $f_1(x,y)$ and $f_2(x,y)$ are same as defined before, and
$$
f_0(x) = \alpha_1 \times (1-e^{-x})^{\frac{\alpha_0^2+\alpha_1^2+\alpha_0 \alpha_1 - \alpha_0}{\alpha_1}} \times
\left (1 - (1-e^{-x})^{\frac{\alpha_0}{\alpha_1}} \right ).
$$
\noindent {\sc Proof:} To prove Theorem 4, we need to show that for all $0 < x,y < \infty$,
$$
F_{X_n,X_{n+1}}(x,y) = \int \int_{B_1} f_1(u,v) du dv + \int \int_{B_2} f_2(u,v) du dv + \int_0^{h(x,y)} f_0(u) |\gamma'(u)|du,  
$$
here for $A(x,y) = \{(u,v); 0 < u \le x, 0 < v \le y\}$, $B_1 = A(x,y) \cup S_1$, $B_2 = A(x,y) \cup S_2$, and
$\ds h(x,y) = \min \left \{x, -\ln \left [ 1 - (1-e^{-y})\frac{\alpha_1}{\alpha_0} \right ] \right \}$.  The first
part, namely
$$
\int \int_{B_1} f_1(u,v) du dv + \int \int_{B_2} f_2(u,v) du dv = p F_a(x,y),
$$
has already shown in Theorem 3.  Therefore, the result is proved if we can show that
$$
\int_0^{h(x,y)} f_0(u) |\gamma'(u)|du = (1-p) F_s(x,y).
$$
Since,
$$
|\gamma'(u)| = \frac{\alpha_0}{\alpha_1} \times \frac{1}{1 - (1-e^{-u})^{\frac{\alpha_0}{\alpha_1}}} \times
(1-e^{-u})^{\frac{\alpha_0}{\alpha_1}-1} \times e^{-u},
$$
\beanno
\int_0^{h(x,y)} f_0(u) |\gamma'(u)|du & = & \alpha_0 \int_0^{h(x,y)} e^{-u} (1-e^{-u})^{\frac{\alpha_0^2+\alpha_1^2+\alpha_0\alpha_1}{\alpha_1}-1} du  \\
& = & \left . (1-p) (1-e^{-u})^{\frac{\alpha_0^2+\alpha_1^2+\alpha_0\alpha_1}{\alpha_1}} \right |_0^{h(x,y)} =
(1-p) F_s(x,y).
\eeanno
\qed

Observe that if $\alpha_0 = \alpha_1 = \alpha$, then (\ref{joint-pdf}) can be written as
$$
f_{X_n,X_{n+1}}(x,y) = \left \{ \begin{array}{ccc}
  f_1(x,y) & \hbox{if} & x < y \\
  f_2(x,y) & \hbox{if} & x > y \\
  f_0(x) & \hbox{if} & x = y,
\end{array}
\right .
$$
here
\beanno
f_1(x,y) & = & 2 \alpha^2 e^{-(x+y)} \left (1 - e^{-x} \right )^{2 \alpha-1} \left (1 - e^{-y} \right )^{\alpha-1} \\
f_2(x,y) & = & 2 \alpha^2 e^{-(x+y)} \left (1 - e^{-x} \right )^{\alpha-1} \left (1 - e^{-y} \right )^{2\alpha-1} \\
f_0(x) & = & \alpha e^{-x} \left (1 - e^{-x} \right )^{3 \alpha-1}.
\eeanno
It can be easily seen that when $\alpha_0 = \alpha_1$, then $(X_n,X_{n+1})$ follows a bivariate generalized exponential
distribution as proposed by Kundu and Gupta \cite{KG:2009}.  Based on the Markovian property of $\{X_n\}$, the
joint PDF of $X_1, \ldots, X_n$ can be written
as
\be
f_{X_1, \ldots, X_n}(x_1, \ldots, x_n) = \frac{\prod_{i=1}^{n-1} f_{X_i,X_{i+1}}(x_i,x_{i+1})}{\prod_{i=2}^n f_{X_i}(x_i)},
\label{jp}
\ee
and this will be useful to develop likelihood inference.  Now we will study the behavior of the maximum
and minimum of a GE process.  Let
$$
Y_n = \max\{X_1, \ldots, X_n\} \ \ \ \ \hbox{and} \ \ \ \ Z_n = \min\{X_1, \ldots, X_n\}.
$$
Then it can be easily seen that for $\alpha = \min\{\alpha_0,\alpha_1\}$,
$$
P(Y_n \le x) = P(X_1 \le x, \ldots, X_n \le x) = (1-e^{-x})^{\alpha_0+\alpha_1+\alpha}
$$
and
$$
P(Z_n \ge x) = P(X_1 \ge x, \ldots, X_n \ge x) = \left (P(X_2 \ge x|X_1 \ge x) \right )^{n-1} P(X_1 \ge x).
$$
Moreover, by simple calculation, it follows that
$$
P(X_2 \ge x|X_1 \ge x) = 1 - \frac{(1-e^{-x})^{\alpha_0+\alpha_1}(1-(1-e^{-x})^{\alpha})}{1-(1-e^{-x})^{\alpha_0+\alpha_1}}.
$$
Now we will discuss about the stopping time.  We define the stopping time as the minimum time so that the process
exceeds a certain level say $L$.  Let us define a new discrete random variable $N$, which denotes the stopping
time, i.e.
$$
\{N = k\} \Leftrightarrow \{X_1 \le L, X_2 \le L, \ldots, X_{k-1} \le L, X_k > L\}.
$$
Here $k$ can take values 1,2, $\ldots$.  Therefore, if $p = (1-e^{-L})$, then 
\beanno
P(N=1) & = & P(X_1 > L) = 1 - P(X_1 \le L) = 1-p^{\alpha_0+\alpha_1} \\
P(N=2) & = & P(X_1 \le L, X_2 > L) = P(X_1 \le L) - P(X_1 \le L, X_2 \le L) =
p^{\alpha_0+\alpha_1} - p^{\alpha_0+\alpha_1+\alpha}  \\
& \vdots & \\
P(N = k) & = & p^{\alpha_0+\alpha_1+(k-2) \alpha} - p^{\alpha_0+\alpha_1+(k-1) \alpha}.
\eeanno
The probability generating function $G_N(s)$ for $N$ becomes
$$
G_N(s) = E(s^N) = \frac{(1-p^{\alpha_0+\alpha_1})z+z^2(p^{\alpha_0+\alpha_1}-p^{\alpha})}{1-p^{\alpha} z}.
$$
Different properties of the stopping time can be obtained from the probability generating function
of $N$.

\section{\sc Maximum Likelihood Estimation}

In this section we consider the maximum likelihood estimation of the unknown parameters based on a random sample of
size $n$, say ${\ve x} = \{x_1, \ldots, x_n\}$, from a GE process.  It is further assumed that the common scale parameter
is also present, i.e. $X_n$ can be written as
\be
X_n = \frac{1}{\lambda} \max \left \{-\ln \left (1 - U_n^{\frac{1}{\alpha_0}} \right ),
-\ln \left (1 - U_{n-1}^{\frac{1}{\alpha_1}} \right ) \right \}.
\ee
It is clear that the $\{X_n\}$ process has two shape parameters $\alpha_0$ and $\alpha_1$, and one scale parameter
$\lambda$.  We consider two cases separately (i) $\alpha_0 = \alpha_1$ and (ii) $\alpha_0 \ne \alpha_1$.
It is observed that when $\alpha_0 = \alpha_1 = \alpha$, the MLEs can
be obtained by solving only one non-linear equation. On the other hand when $\alpha_0 \ne \alpha_1$, the MLEs
are obtained by solving a two-dimensional optimization problem. 

\subsection{\sc Case I: $\alpha_0 = \alpha_1 = \alpha$}

Note that in presence of the scale parameter, the joint density function of $X_n$ and $X_{n+1}$ is (\ref{joint-pdf}),
where
\bea
f_1(x,y) & = & 2 \alpha^2 \lambda^2 e^{-\lambda(x+y)} \left (1 - e^{-\lambda x} \right )^{2 \alpha-1}
\left (1 - e^{-\lambda y} \right )^{\alpha-1}  \nonumber \\
f_2(x,y) & = & 2 \alpha^2 \lambda^2 e^{-\lambda(x+y)} \left (1 - e^{-\lambda x} \right )^{\alpha-1}
\left (1 - e^{-\lambda y} \right )^{2\alpha-1}  \nonumber \\
f_0(x) & = & \alpha \lambda e^{-\lambda x} \left (1 - e^{-\lambda x} \right )^{3 \alpha-1}.  \label{joint-pdf-eqs}
\eea
Let us use the following notations:
$$
A_1 = \{i; 1 \le i \le n-1, x_i < x_{i+1}\}, \ \
A_2 = \{i; 1 \le i \le n-1, x_i > x_{i+1}\}, \ \
$$
$$
A_0 = \{i; 1 \le i \le n-1, x_i = x_{i+1}\}, \ \
$$
and
$$
|A_0| = n_0, |A_1| = n_1, |A_2| = n_2.
$$
Clearly, $n_0+n_1+n_2 = n-1$.  Moreover, we also define $f_{X_i}(x;\alpha,\lambda)$ and
$f_{X_i|X_{i-1}=y}(x;\alpha,\lambda)$ as the density function of $X_i$ and the conditional density function of $X_i$
given $X_{i-1}=y$, respectively.  Now based on the observations, the log-likelihood function of the observed data
becomes, see (\ref{jp}),
\beanno
l(\alpha,\lambda|Data) & = & \ln f_{X_1}(x_1;\alpha,\lambda)
+ \sum_{i=2}^n \ln f_{X_i|X_{i-1} = x_{i-1}}(x_i;\alpha,\lambda)  \nonumber \\
& = & \sum_{i=1}^{n-1} \ln f_{X_i,X_{i+1}}(x_i,x_{i+1};\alpha,\lambda) - \sum_{i=2}^{n-1} \ln f_{X_i}(x_i;\alpha,\lambda)
\nonumber \\
& = & \sum_{i \in I_0 \cup I_i \cup I_2} \ln f_{X_i,X_{i+1}}(x_i,x_{i+1};\alpha,\lambda) - \sum_{i=2}^{n-1} \ln f_{X_i}(x_i;\alpha,\lambda)
\nonumber \\
& = & c + (n_1+n_2+2) \ln \alpha + \alpha g_1(\lambda, {\ve x}) + (n_1+n_2+2) \ln \lambda - \lambda g_2({\ve x})
+ g_3 (\lambda, {\ve x}).
\eeanno
Here $c$ is a constant, and
\beanno
g_1(\lambda, {\ve x}) & = & 2 \sum_{i \in A_1} \ln (1 - e^{-\lambda x_i} ) + \sum_{i \in A_1} \ln (1 - e^{-\lambda x_{i+1}} )
+ \sum_{i \in A_2} \ln (1 - e^{-\lambda x_i} ) + 2 \sum_{i \in A_2} \ln (1 - e^{-\lambda x_{i+1}} ) \\
& & - \sum_{i=2}^{n-1} \ln (1 - e^{-\lambda x_i} )  \\
g_2({\ve x}) & = & \left [ \sum_{i \in A_1 \cup A_2}(x_i + x_{i+1}) + \sum_{A_0} x_i - \sum_{i=2}^{n-1} x_i \right ]  \\
g_3(\lambda,{\ve x}) &= & - \sum_{i \in A_1 \cup A_2} \ln (1 - e^{-\lambda x_i} ) -
  \sum_{i \in A_1 \cup A_2} \ln (1 - e^{-\lambda x_{i+1}} ) - \sum_{i \in A_0} \ln (1 - e^{-\lambda x_i} ) +
  \sum_{i = 2}^{n-1} \ln (1 - e^{-\lambda x_i} ).
\eeanno
Hence, for a given $\lambda$, the MLE of $\alpha$, say, $\widehat{\alpha}(\lambda)$ can be obtained as
\be
\widehat{\alpha}(\lambda) = -\frac{n_1+n_2+2}{g_1(\lambda, {\ve x})},    \label{mle-al}
\ee
and the MLE of $\lambda$ can be obtained by maximizing
$$
h(\lambda) = -(n_1+n_2+2) \ln (g_1(\lambda,{\ve x})) + (n_1+n_2+2) \ln \lambda - \lambda g_2({\ve x})
+ g_3 (\lambda, {\ve x}).
$$
{\color{black} The maximization of $h(\lambda)$ involves solving a one-dimensional optimization problem.  We can
  use bisection method to compute $\widehat{\lambda}$, the maximum of $h(\lambda)$.  Once $\widehat{\lambda}$ is
  obtained then $\widehat{\alpha}$ can be obtained as $\widehat{\alpha}(\widehat{\lambda})$ from (\ref{mle-al}).
  Due to complicated
  nature of $h(\lambda)$, it is not possible to prove that it is an unimodal function.  We propose to plot
  $h(\lambda)$ as a function of $\lambda$, to get an idea about $\widehat{\lambda}$.  The details have been
  illustrated in the Synthetic Experiments and Data Analysis sections.}

\subsection{\sc Case II: $\alpha_0 \ne \alpha_1$}

Now we will consider the MLEs of the unknown parameters, when $\alpha_0 \ne \alpha_1$.  Therefore, in this case
we have three unknown parameters.  In this case the joint PDF of $X_n$ and $X_{n+1}$ in presence of the scale
parameter $\lambda$ is (\ref{joint-pdf}),
where
\beanno
f_1(x,y) & = & \alpha_0 (\alpha_0+\alpha_1) \lambda^2 e^{-\lambda(x+y)} \left (1 - e^{-\lambda x} \right )^{\alpha_0
  +\alpha_1-1} \left (1 - e^{-\lambda y} \right )^{\alpha_0-1}  \nonumber \\
f_2(x,y) & = & \alpha_1 (\alpha_0+\alpha_1) \lambda^2 e^{-\lambda(x+y)} \left (1 - e^{-\lambda x} \right )^{\alpha_1-1}
\left (1 - e^{-\lambda y} \right )^{\alpha_0+\alpha_1-1}  \nonumber \\
f_0(x) & = & \alpha_1 \lambda \left (1 - e^{-\lambda x} \right )^{\frac{\alpha_0^2+\alpha_1^2+\alpha_0 \alpha_1 -\alpha_0}{\alpha_1}}\left (1 - \left (1 - e^{-\lambda x} \right )^{\frac{\alpha_0}{\alpha_1}} \right ) .  
\eeanno
We reparametrize as $(\alpha_0,\alpha_1,\lambda)$ to $(\gamma,\alpha_1,\lambda)$, where
$\ds \gamma = \frac{\alpha_0}{\alpha_1}$.  We use the following notations:
\beanno
A_1(\gamma) & = & \left \{i; 1 \le i \le n-1, (1-e^{-\lambda x_i}) < (1-e^{-\lambda x_{i+1}})^{\gamma} \right \},   \\
A_2(\gamma) & = & \left \{i; 1 \le i \le n-1, (1-e^{-\lambda x_i}) > (1-e^{-\lambda x_{i+1}})^{\gamma} \right \},   \\
A_0(\gamma) & = & \left \{i; 1 \le i \le n-1, (1-e^{-\lambda x_i}) = (1-e^{-\lambda x_{i+1}})^{\gamma} \right \}, 
\eeanno
and $n_1(\gamma) = |A_1(\gamma)|$, $n_2(\gamma) = A_2(\gamma)|$, $n_0(\gamma) = |A_0(\gamma)|$.
Therefore, based on the above notations, the log-likelihood function becomes
\beanno
l(\gamma,\alpha_1,\lambda|{\ve x}) & = &
\ln f_{X_1}(x_1;\gamma,\alpha_1,\lambda)
+ \sum_{i=2}^n \ln f_{X_i|X_{i-1} = x_{i-1}}(x_i;\gamma,\alpha_1,\lambda)  \nonumber \\
& = & \sum_{i=1}^{n-1} \ln f_{X_i,X_{i+1}}(x_i,x_{i+1};\gamma,\alpha_1,\lambda) -
\sum_{i=2}^{n-1} \ln f_{X_i}(x_i;\gamma,\alpha_1,\lambda)  \nonumber \\
& = & (2n_1(\gamma)+2n_2(\gamma)+n_0(\gamma)) \ln \alpha_1 + \alpha_1 h_1(\lambda,\gamma,{\ve x}) + \\
&  & (2n_1(\gamma)+2n_2(\gamma)+n_0(\gamma)) \ln \lambda - \lambda  \sum_{i \in A_1(\gamma) \cup A_2(\gamma)} (x_i+x_{i+1})
+  \\
& & n_1(\gamma)(\ln \gamma + \ln (1+\gamma)) + n_2(\gamma) \ln (1+\gamma) + \\
& & \sum_{i \in A_0(\gamma)} \ln (1 - (1-e^{-\lambda x_i})^{\gamma}) - \sum_{i \in A_1(\gamma) \cup A_2(\gamma)} \ln (1-e^{-\lambda x_i})
- \\
& & \sum_{i \in A_1(\gamma) \cup A_2(\gamma)} \ln (1-e^{-\lambda x_{i+1}}) - \sum_{i \in A_0(\gamma)} \ln (1-e^{-\lambda x_i})
\eeanno
where
\beanno
h_1(\lambda,\gamma,{\ve x}) & = & (1+\gamma) \sum_{i \in A_1(\gamma)} \ln (1-e^{-\lambda x_i}) +
\gamma \sum_{i \in A_1(\gamma)} \ln (1-e^{-\lambda x_{i+1}}) + \nonumber \\
&  &  (1+\gamma) \sum_{i \in A_2(\gamma)} \ln (1-e^{-\lambda x_{i+1}}) + \gamma \sum_{i \in A_2(\gamma)} \ln (1-e^{-\lambda x_i}) +\\
& &  (1+\gamma+\gamma^2) \sum_{i \in A_0(\gamma)} \ln (1-e^{-\lambda x_i}). 
\eeanno
For fixed $\gamma$ and $\lambda$, the MLE of $\alpha_1$, say $\widehat{\alpha}_1(\gamma,\lambda)$ can be obtained as
\be
\widehat{\alpha}_1(\gamma,\lambda) = - \frac{2n_1(\gamma)+2n_2(\gamma)+n_0(\gamma)}{h_1(\lambda,\gamma,{\ve x})}.
\label{mle-ne-al}
\ee
Hence, the MLEs of $\gamma$ and $\lambda$ can be obtained by maximizing numerically $\ds l(\gamma,\widehat{\alpha}_1(\gamma,\lambda), \lambda)$ with respect to $\gamma$ and $\lambda$.
{\color{black} Note that it is a two-dimensional optimization problem.  Newton-Raphson or some
  iterative methods may be used to compute $\widehat{\gamma}$ and $\widehat{\lambda}$, the MLEs of $\gamma$ and $\lambda$,
  respectively.  Once $\widehat{\gamma}$ and $\widehat{\lambda}$ are obtained, the MLE of $\alpha$ can be obtained
  as $\widehat{\alpha}(\widehat{\gamma},\widehat{\lambda})$ from (\ref{mle-ne-al}).  One needs a starting values
  to start any iterative process.  In this respect, we suggest to use the contour plot of the profile log-likelihood
  function $\ds l(\gamma,\widehat{\alpha}_1(\gamma,\lambda), \lambda)$.  The details will be illustrated  in the
  Synthetic Experiments and Data Analysis sections.}

{\color{black}
\section{\sc Synthetic Experiments}

In this section we provide the analyses of two synthetic data sets.  Two data sets have been simulated:}
(i) $\alpha_0 = \alpha_1 = \alpha$, (ii) $\alpha_0 \ne \alpha_1$.

{\color{black}
\subsection{\sc Synthetic Data Set 1:}}

\begin{figure}[h]
\begin{center}
\includegraphics[height=6.0cm,width=7.0cm]{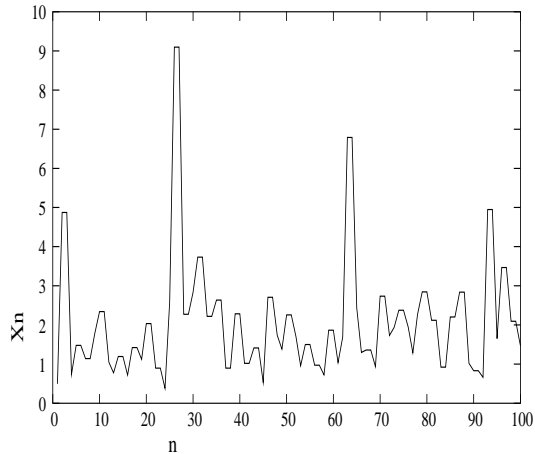}

\caption {Generated $X_n$, when $\alpha_0 = \alpha_1 = \alpha$ = 2.0 and $\lambda$ = 1.}   \label{data-1}
\end{center}
\end{figure}

\begin{figure}[h]
\begin{center}
\includegraphics[height=6.0cm,width=7.0cm]{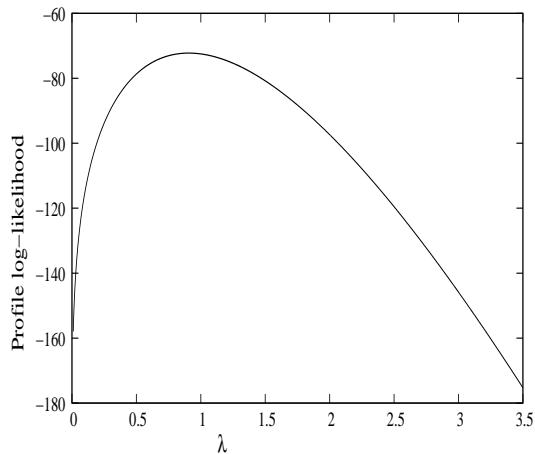}

\caption {The profile log-likelihood of $\lambda$.}   \label{ge-data-prof}
\end{center}
\end{figure}

In this case we have generated the data set of size $n$ = 100 with the following parameters:
$$
\alpha_0 = \alpha_1 = \alpha = 2.0 \ \ \ \ \hbox{and} \ \ \ \  \lambda =  1.0.
$$
The generated $\{x_1, \ldots, x_{100}\}$ has been plotted in Figure \ref{data-1}.  We first computed the MLE of $\lambda$
by maximizing the profile log-likelihood function $h(\lambda)$ as mentioned in the previous section.  The
profile log-likelihood function $h(\lambda)$ has been plotted in Figure \ref{ge-data-prof}.  It is an unimodal
function.  Therefore, MLEs are unique in this case.  {\color{black} It clearly gives an idea that the MLE of $\lambda$
  lies between 0.5 and 1.5.  We start our bisection method with these two boundaries, and 
the MLE of
$\lambda$ has been obtained as $\widehat{\lambda}$
= 0.9058}.  Based on $\widehat{\lambda}$, the MLE of $\alpha$ can be obtained as $\widehat{\alpha}$  =1.5164.
We have used parametric bootstrap method to compute the 95\% confidence intervals of $\alpha$ and $\lambda$ and
they are $(0.7132,1.1015)$ and $(0.9054,2.1141)$, respectively.

{\color{black}
\subsection {\sc Synthetic Data Set 2:}}

\begin{figure}[h]
\begin{center}
\includegraphics[height=6.0cm,width=7.0cm]{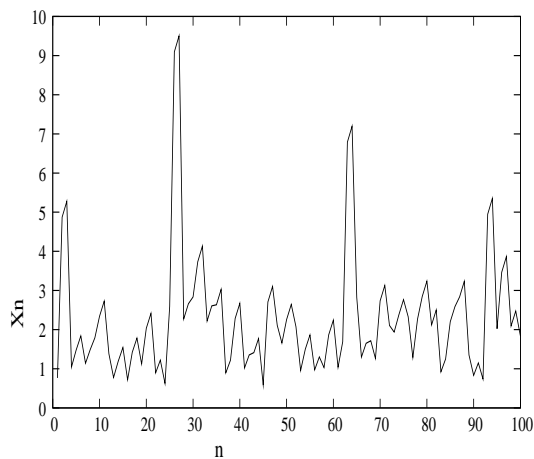}

\caption {Generated $X_n$, when $\alpha_0$ = 2.0, $\alpha_1$ = 3.0 and $\lambda$ = 1.}   \label{data-2}
\end{center}
\end{figure}

In this case a data set of size $n$ = 100, has been generated using the
following parameters:
$$
\alpha_0 = 2.0, \ \ \ \ \alpha_1 = 3.0, \ \ \ \ \hbox{and} \ \ \ \  \lambda =  1.0.
$$
The data set has $\{x_1, \ldots, x_{100}\}$ has been plotted in Figure \ref{data-2}. 

\begin{figure}[h]
\begin{center}
\includegraphics[height=6.0cm,width=7.0cm]{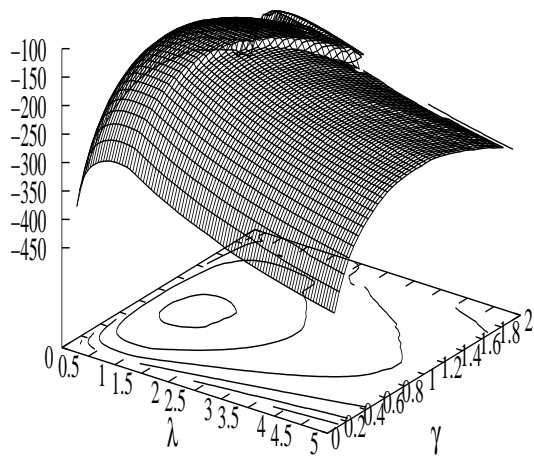}

\caption {The contour plot of the profile log-likelihood of $\lambda$ and $\gamma$.}   \label{ge-data-contour}
\end{center}
\end{figure}

Now we would like to compute the MLEs of the $\alpha_1$, $\gamma$ and $\lambda$, and they can be obtained
by maximizing first the profile log-likelihood function $h_1(\lambda,\gamma,{\ve x})$ as defined in Section
4.2.  We provide the contour lot of $h_1(\lambda,\gamma,{\ve x})$ in Figure \ref{ge-data-contour}.
It indicates that the profile log-likelihood function of $\lambda$ and $\gamma$ is an unimodal function, hence
the MLEs are unique.  {\color{black} The contour plot provides a good idea about the initial guesses of $\lambda$
  and $\gamma$.  We have started the iterative process with $\lambda$ = 1 and $\gamma$ = 0.6.  The iteration
  stops at $\widehat{\lambda}$ = 0.8699 and $\widehat{\gamma}$ = 0.8500}.  Based on these, 
the MLEs of $\alpha_0$, $\alpha_1$ and $\lambda$ become $\widehat{\alpha}_0$ = 2.1338,
$\widehat{\alpha}_1$ = 2.5103 and $\widehat{\lambda}$ = 0.8699.  In this case based on the parametric bootstrap
the 95\% confidence intervals for $\alpha_0$, $\alpha_1$ and $\lambda$ are (1.5431,2.8342), (1.8775,3.2312)
and (0.6754,1.1231), respectively.

\section{\sc Gold Price Data Analysis}

\begin{figure}[t]
\begin{center}
\includegraphics[height=6.0cm,width=7.0cm]{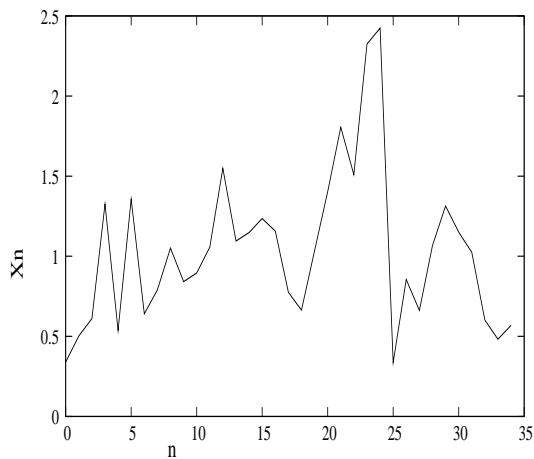}

\caption {Gold price in Indian market for 45 days starting from October 06, 2020.}   \label{gold-price}
\end{center}
\end{figure}

\begin{figure}[t]
\begin{center}
\includegraphics[height=6.0cm,width=7.0cm]{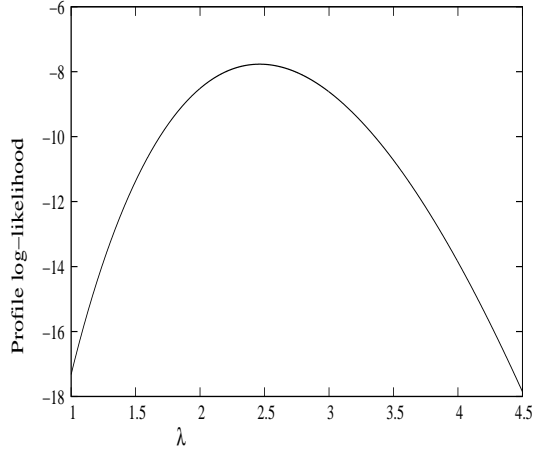}

\caption {The profile log-likelihood function of $\lambda$ for Gold price data.}   \label{gold-profile}
\end{center}
\end{figure}

{\color{black} In this section we present the analysis of gold-price
  data  based on the proposed GE  process to see how the proposed model and methods can be used in practice.  
This is a real data set of gold price per gram in Indian Rupees in Indian market of 45 days starting from
October 06, 2020, and it has been obtained from the website as follows: 
  https://www.bullion-rates.com/gold/INR-history.htm. 
There is no trading during the 
weekends and holidays, hence we have data for 35 days.  The minimum and maximum values were  Rs. 4230.02 and
Rs. 4642.32, respectively.  We have scaled the data set by subtracting 4200 and divided by 100, to each data points.  The scaled data set
has been}
plotted in Figure \ref{gold-price}.
First we compute the MLEs of $\alpha$ and $\lambda$ based on the assumption $\alpha_0  = \alpha_1 = \alpha$.
The profile log-likelihood function of $\lambda$ has been plotted in Figure \ref{gold-profile}.
It is an unimodal function, hence the MLEs are unique.  The MLEs of $\alpha$ and $\lambda$ are
$\widehat{\lambda}$ = 2.4620 and $\widehat{\alpha}$ = 3.3498.  The associated log-likelihood value is -11.4732,
95\% confidence intervals of $\lambda$ and $\alpha$ are (1.9854,2.6831) and (2.9552,3.7454), respectively.

\begin{figure}[t]
\begin{center}
\includegraphics[height=6.0cm,width=7.0cm]{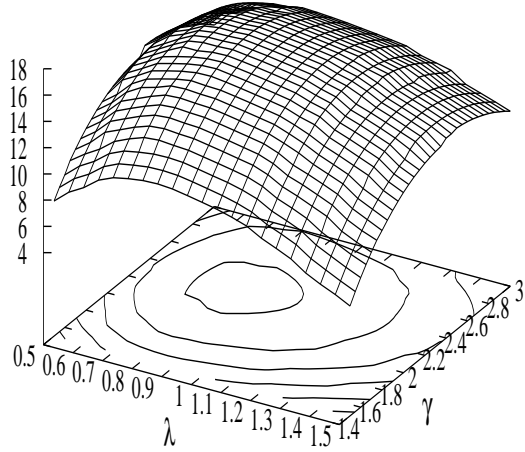}

\caption {The contour plot of $\gamma$ and $\lambda$ for the gold-price log-likelihood function.}   \label{gold-contour}
\end{center}
\end{figure}

Based on the assumption $\alpha_0 \ne \alpha_1$, the MLEs of $\lambda$, $\alpha_0$ and $\alpha_1$ are
$\widehat{\lambda}$ = 2.3449, $\widehat{\alpha}_0$ = 3.5312, $\widehat{\alpha}_1$ = 4.2684.  The associated
log-likelihood value is -9.0123 and 95\% confidence intervals of  $\lambda$, $\alpha_0$ and $\alpha_1$ are
(1.9756,2.7016), (2.9625,3.7523), (3.9598,4.6734), respectively.  The contour plot of $\lambda$ and $\gamma$
is provided in Figure \ref{gold-contour}.
From the
contour plot of $\gamma$ and $\lambda$, it is clear that the MLEs exist and they are unique.

Now one natural question is how to show that GE process fits the gold price data.  We still do not have a
proper goodness of fit test, but we have tried the following measures which ensures at least that it does
not violate some of the sufficient conditions.  If $\{X_n\}$ is a GE process, then
$\{X_{2n-1}; n = 1,2, \ldots \}$, will be i.i.d. GE random variables, and similarity,
$\{X_{2n}; n = 1,2, \ldots\}$ will be also i.i.d. GE random variables.  {\color{black} Now we would like to test the following: first we
  would like to test whether $\{x_1,x_3, \ldots, x_{35}\}$ are independently distributed or not, and then test whether
  they follow GE distribution or not.  The same we want to do for $\{x_2, x_4, \ldots, x_{34}\}$ also.  To test
  whether they are independent or not we have used run test, and for testing whether they follow GE distributions or
  not we have used Kolmogorv-Smirnov (KS) test.}

We fit GE($\beta, \theta$) to
$\{x_1, x_3, \ldots, x_{35}\}$.  The MLEs of $\beta$ and $\theta$ are $\widehat{\beta}$ = 7.0863 and
$\widehat{\theta}$ = 2.5866, respectively.  
The Kolmogorov-Smirnov (KS) distance between the empirical
cumulative distribution function (CDF) and the fitted CDF is 0.1437 and the corresponding $p$ value is
0.8513.  We further perform the test of independence of $\{x_1, x_3, \ldots, x_{35}\}$ based on run test, and the
corresponding $p$ value becomes 0.15.  We have done the same procedure for $\{x_2, x_4, \ldots, x_{34}\}$
also.  In this case $\widehat{\beta}$ = 6.6324 and $\widehat{\theta}$ = 2.3979.  The KS distance and the
corresponding $p$ values are 0.1326 and 0.9261, respectively.  The $p$ value based on run test is 0.45.
Hence, we cannot reject the null hypothesis that $\{x_1, x_3, \ldots, x_{35}\}$ is a random sample from
a GE distribution, and the same for $\{x_2, x_4, \ldots, x_{34}\}$.  

We have further computed the first and second order autocorrelations of the data set and they are 0.0915
and 0.0402, respectively.  We have obtained the distribution of the first order and second order
autocorrelations for GE process based on simulations.  The upper 90\% percentile points of the first and
second order autocorrelations are 0.1621 and
0.0504, respectively.  Therefore, based on the observed first and second order autocorrelations, we cannot
reject the hypothesis
that the data are coming from a GE process and based on the log-likelihood values, we cannot reject the
hypothesis that $\alpha_0 = \alpha_1$.  Hence, we conclude that GE process with two equal shape parameters,
fits the gold-price data well.

\section{\sc Conclusions}

{\color{black} In this paper we propose a new discrete time and continuous state space stationary process, and
  we named it as a GE process.  It is called a stationary GE process as the marginals are GE distributions and
  it is a stationary process.  The distinct feature of this proposed process is that the joint distribution of
  $X_n$ and $X_{n+1}$ is a singular distribution, due to this it can be used if there are some
  ties in the consecutive data points with positive probability.  The existing Weibull or
  gamma processes do not have this feature.
 
The proposed GE process}
can be easily extended to a more general class of
proportional reversed hazard (PRH) process, i.e. for a class of lifetime distribution functions which can be
represented as follows:
$$
F(t; \alpha) = \left (F_0(t) \right )^{\alpha}.
$$
Here $F_0(t)$ is an absolutely continuous distribution, and $F_0(t) = 0$, for
$t \le 0$, see for example Kundu and Gupta \cite{KG:2010}.  Now we can define PRH process as follows.
$$
X_n = \max\{F_0^{-1}(U_n^{\frac{1}{\alpha_0}}),F_0^{-1}(U_{n-1}^{\frac{1}{\alpha_1}})\},
$$
where $\{U_0, U_1, \ldots, \}$ is a sequence of i.i.d. Uniform (0,1) random variables, $\alpha_0 > 0$
and $\alpha_1 > 0$.
Most of the results what we have developed for the GE process, can be extended for the PRH process also.
It will be interesting to develop proper inferential procedure and some model selection criteria for data
analysis purposes.  More work is
needed along that direction.

{\color{black}
\section*{\sc Acknowledgements:} The authors would like to thank the unknown reviewers for making constructive
suggestions which have helped to improve the earlier version of the manuscript significantly.

\section*{\sc Conflict of Interest and Funding Statements:} The author does not have any conflict of interest.  The
author did not receive any funding from any source in preparation of this manuscript.
}

\end{document}